\documentclass[prd,aps,showpacs,nofootinbib,eqsecnum]{revtex4-2}
\usepackage{graphicx,color,amsmath,amsxtra}
\usepackage{amssymb}
\usepackage[english]{babel}
\usepackage{amsfonts}
\allowdisplaybreaks[4]

\newcommand{\beq}{\begin{equation}}
	\newcommand{\eeq}{\end{equation}}
\newcommand{\bea}{\begin{eqnarray}}
	\newcommand{\eea}{\end{eqnarray}}
\newcommand{\bseq}{\begin{subequations}}
	\newcommand{\eseq}{\end{subequations}}

\newcommand{\rf}[1]{(\ref{#1})}

\newcommand{\rma}{\textrm{a}}
\newcommand{\coupl}{\eta} 
\newcommand{\calH}{\mathcal{H}} 
\newcommand{\ct}{\tau} 

\begin{document}

\title{Cosmological perturbations in the theory of gravity with non-minimal derivative coupling. I. Modes of perturbations} 
	
\author{R. I. Kamalitdinov}
\author{S. V. Sushkov}
\email{sergey\_sushkov@mail.ru}
\address{Institute of Physics, Kazan Federal University,
Kremlevskaya 16a, Kazan 420111, Russia}

\begin{abstract}
We consider perturbations in the isotropic and homogeneous cosmological model with the spatially flat Friedmann–Lema\^{i}tre–Robertson–Walker metric in the framework of the theory of gravity with non-minimal derivative coupling. 
The Lagrangian of the theory contains the coupling term $\coupl G^{\mu\nu}\nabla_\mu\phi \nabla_\nu\phi$ and represents the particular example of a general Horndeski Lagrangian, which results in second-order field equations. 
It is known that the non-minimal derivative coupling {\em crucially} changes scenarios of the Universe evolution on early times. 
In particular, the $\coupl$-term is dominating on early times and leads to
a primary quasi-de Sitter (inflationary) stage which needs no fine-tuned potential.
On late times the influence of non-minimal derivative coupling on the Universe evolution completely disappears, and this naturally leads to the transition to the standard cosmological evolution (post-inflationary stage).  
We have derived a complete set of equations which describe an evolution of scalar, vector and tensor modes of perturbations. All modes are analyzed analytically in two asymptotic cases, and then we construct exact numerical solutions which describe an entire evolution of the modes.
We show that all modes, including vector ones, are amplified in the quasi-de Sitter (inflationary) stage, and such the behavior is cardinally distinct from that in Friedmann cosmology.
\end{abstract}

\maketitle

\section{Introduction}
In recent decades the observational cosmology has been going through the period of the rapid growth. Precise measurements of the Cosmic Microwave Background (CMB) radiation \cite{CMB1,CMB2}, systematic observations of nearby and distant Type Ia supernovae (SNe Ia) \cite{SN1,SN2,SN3,SN4,SN5,SN6,SN7,SN8}, study of  baryon acoustic oscillations \cite{BAO1,BAO2,BAO3,BAO4,BAO5,BAO6,BAO7}, mapping the large-scale structure of the Universe, microlensing observations, and many other remarkable results (see, for example, the review \cite{Observations}) have essentially expanded our knowledge about the Universe. Amazing discoveries, such as the accelerating expansion of the Universe and the dark matter evidence, have set new serious challenges before theoretical cosmology faced the necessity of radical modification of the standard model having successfully been exploited for a long time. Now, any viable cosmological model has to be able to describe several qualitatively different epoches of the Universe evolution, including the primary inflation, the matter-dominated stage, and the present acceleration (or secondary inflation). Moreover, it should also explain a mechanism providing an epoch change. These challenges have prompted many speculations mostly based on phenomenological ideas which involve new dynamical sources of gravity that act as dark energy, and/or various modifications to general relativity. To date, many different versions of modified or extended theories of gravity have been proposed (see surveys 
\cite{Review_Salvatore:2011, Review_Clifton_etal, Review_ModGrav:2013, Review_Berti_etal, Review_Nojiri:2017, Review_Langlois:2019} and references therein).

One of the modified theories of gravity intensively studied today is the Horndeski theory \cite{Horndeski} derived in the 1970s as an attempt to obtain the most general action for a scalar-tensor theory with a single scalar degree of freedom and second-order field equations. In 2011 Horndeski gravity has been rediscovered in the context of generalized Galileon theories \cite{Kobayashi:2011}, and since the interest in this model has only growing.\footnote{The literature dedicated to various aspects of Horndeski gravity is very vast, and its survey lays out of the scope of this work. The reader interesting in this topic can find some references in the already mentioned surveys \cite{Review_Berti_etal, Review_Clifton_etal}.}
The important subclass of the Horndeski theory is represented by theories with the non-minimal derivative coupling. In these models the Lagrangian contains an additional term $\coupl G^{\mu\nu}\nabla_\mu\phi \nabla_\nu\phi$, which directly provides non-minimal coupling between derivatives of the scalar field and the spacetime curvature expressed in terms of the Einstein tensor $G_{\mu\nu}$, and $\coupl$ is the coupling parameter with dimension of ({\em length})$^2$. 

The non-minimal derivative coupling leads to very interesting and important consequences, both astrophysical and cosmological. 
In particular, black holes \cite{Rinaldi:2012, Minamitsuji:2013, Anabalon:2014, Babichev:2014, Kobayashi:2014, Babichev:2015}, wormholes \cite{Sushkov:2012b, Sushkov:2014}, and neutron stars \cite{Rinaldi:2015, Rinaldi:2016, Silva:2016, Maselli:2016, Eickhoff:2018, KasSus:2023, Kashargin:2025iao} have been widely explored within this theory. 
Cosmological models with the non-minimal derivative coupling have been also intensively studied in the literature (see Refs.
\cite{Sushkov:2009, SarSus:2010, Sushkov:2012, SkuSusTop:2013, MatSus:2015, StaSusVol:2016, StaSusVol:2019, Galeev_etal:2021, Sushkov:2023aya} and references therein). 
%
It was shown that the non-minimal derivative coupling {\em crucially} changes scenarios of the Universe evolution on early times.
In particular, the $\coupl$-terms turn out to be dominating on early times and totally screen the influence of the cosmological constant and the material filling. The most important consequences of non-minimal derivative coupling are an primary {\em kinetic} inflation which needs no fine-tuned potential \cite{Sushkov:2012,Sushkov:2023aya}, the existence of bounce scenario \cite{Sushkov:2025udy}, an anisotropy screening on early times of the Universe evolution \cite{StaSusVol:2016, StaSusVol:2019}. At the same time, on late times the influence of non-minimal derivative coupling on the Universe evolution completely disappears, and this naturally leads to the transition to the standard cosmological evolution. 

Any cosmological theory cannot be restricted by considering only isotropic and homogeneous models. It is also necessary  to study, at least, linear perturbations of such the models. Cosmological perturbations in the Horndeski theory and its particular subclasses, and also in some extensions of this theory    
have been discussed in Refs. \cite{Gleyzes:2013,Bellini:2014,Germani:2015,Akama:2018,Ageeva:2022,Mironov:2023,Ahmedov:2023,Santos:2024,deBoe:2025,Dent:2013}. 
In Ref. \cite{Gleyzes:2013} the authors proposes a minimal description of single field dark energy/modified gravity within the effective field theory formalism for cosmological perturbations, which encompasses most existing models. As
an illustration, they study the Horndeski theory and derive the set of equations for  linear scalar fluctuations.
See also Refs. \cite{Dent:2013} and \cite{Bellini:2014}, where the equations of motion for linear scalar perturbations was given in the Newtonian and synchronous gauges, respectively. 
In Ref. \cite{Germani:2015} the authors also focuses only on scalar perturbations and argue that in the Horndeski theory the Newtonian potential can grow to potentially dangerously large values during the post-inflationary phase. This
amplification can potentially violate the linearity of the system, and so the inflationary predictions would be completely lost.
In Ref. \cite{Ageeva:2022} the authors construct a concrete model of Horndeski bounce (without the $G_5$-term providing the non-minimal derivative coupling) with strong gravity in the past, study cosmological perturbations in this model, and 
show that it is possible to generate perturbations in a controllable way,
i.e. in the regime where the background evolution and perturbations are legitimately
described within classical field theory and weakly coupled quantum theory.
In Ref. \cite{Mironov:2023} the perturbations above Bianchi I type background in the the Horndeski theory are studied.
In Refs. \cite{Santos:2024,deBoe:2025} scalar and tensor perturbations are explored in an extended version of the Horndeski theory, when the Horndeski Lagrangian is equipped with two additional dilaton fields.
In Ref. \cite{Ahmedov:2023} cosmological perturbations in the teleparallel analog of Horndeski gravity was considered.
As well let us mention the work \cite{Darabi:2013}, which is, in our opinion, erroneous. Actually, in Ref. \cite{Darabi:2013} the authors study scalar perturbations in the theory of gravity with non-minimal derivative using the Newtonian gauge and supposing \textit{a priory} that two scalar perturbations, $\Psi$ (the Newtonian potential) and $\Phi$ satisfy the relation $\Psi+\Phi=0$. Generally, this is not true for the Horndeski theory.

Here we would like to notice that investigations presented in Refs. \cite{Gleyzes:2013,Bellini:2014,Germani:2015,Akama:2018,Ageeva:2022,Mironov:2023,Ahmedov:2023,Santos:2024,deBoe:2025,Dent:2013,Darabi:2013} are mainly based on qualitative analysis of perturbations. That is, one constructs quadratic actions for perturbations (scalar or tensor) and analyses effective coefficients standing before `kinetic' and `potential' parts of a perturbed Lagrangian. In our knowledge, in the literature there is no systematic and complete analysis of an exact behavior of perturbation modes. Moreover, typically the analysis of perturbations in the Horndeski theory is only restricted to scalar and tensor perturbations, because  usually one supposes that vector perturbations are decayed during all eras of the Universe evolution. This supposition is based on the analogy with general relativity, where it is true. However, in this work we will see that this is not true for the theory of gravity with non-minimal derivative coupling. 

Our aim is to perform a systematic and complete perturbation analysis in the theory of gravity with non-minimal derivative coupling. The results of analysis will be represented in two papers. This one is first. Here we derive the complete set of equations of motion for scalar, tensor and vector modes of perturbations, find their asymptotic and exact numerical solutions and analyzed results obtained. The second paper will be devoted to constructing and analyzing a power spectrum of perturbations. 

The paper is organized as follows. In Section II we briefly describe the theory of gravity with non-minimal coupling. In Section III isotropic and homogeneous cosmological solutions are considered, which are an unperturbed background of the model. In Section IV we construct equations of motion for scalar, tensor and vector modes of perturbations and find their asymptotic and exact numerical solutions. In Section V we discuss results obtained and make the concluding remarks. 

\section{Field equations}
We will consider the theory of gravity with non-minimal derivative coupling with the action given as follows\footnote{We use the units $\hbar=c=G=1$, so that $M_{\text{Pl}}^2=(8\pi)^{-1}$, where $M_{\text{Pl}}$ is the Planck mass.} 
\beq\label{action}
S=\frac12\int d^4x\sqrt{-g}\left[ \frac{1}{8\pi}R
-\big(g^{\mu\nu}+\coupl G^{\mu\nu} \big) \nabla_{\mu}\phi\nabla_{\nu}\phi
\right] +S^{(m)},
\eeq
where $R$ and $G_{\mu\nu}$ are the scalar curvature and the Einstein tensor, respectively, 
$\coupl$ is the coupling parameter with dimension of ({\em length})$^2$, 
and $S^{(m)}$ is the action for ordinary matter fields, supposed to be minimally coupled to gravity in the usual way. Since we will be mostly interesting in the early-time Universe dynamics driven purely by the scalar field, further we will suppose that $S^{(m)}\equiv 0$, i.e. ordinary matter is absent.

Varying the action with respect to $g_{\mu\nu}$ and $\phi$ gives the field
equations, respectively:
\bseq\label{fieldeq}
\bea
\label{eineq}
&& G_{\mu\nu}=
8\pi\big[
T^{(\phi)}_{\mu\nu} +\coupl \Theta_{\mu\nu}\big], \\
\label{eqmo}
&&[g^{\mu\nu}+\coupl G^{\mu\nu}]\nabla_{\mu}\nabla_\nu\phi=0,
\eea
\eseq
where 
\bea \label{T}
T^{(\phi)}_{\mu\nu}&=&\nabla_\mu\phi\nabla_\nu\phi-
{\textstyle\frac12}g_{\mu\nu}(\nabla\phi)^2, \\
\Theta_{\mu\nu}&=&-{\textstyle\frac12}\nabla_\mu\phi\,\nabla_\nu\phi\,R
+2\nabla_\alpha\phi\,\nabla_{(\mu}\phi R^\alpha_{\nu)}
+\nabla^\alpha\phi\,\nabla^\beta\phi\,R_{\mu\alpha\nu\beta}
\nonumber\\
&&
+\nabla_\mu\nabla^\alpha\phi\,\nabla_\nu\nabla_\alpha\phi
-\nabla_\mu\nabla_\nu\phi\,\square\phi
-{\textstyle\frac12}(\nabla\phi)^2 G_{\mu\nu}
\nonumber \\
&&
+g_{\mu\nu}\big[-{\textstyle\frac12}\nabla^\alpha\nabla^\beta\phi\,
\nabla_\alpha\nabla_\beta\phi
+{\textstyle\frac12}(\square\phi)^2
-\nabla_\alpha\phi\,\nabla_\beta\phi\,R^{\alpha\beta}
\big]. 
\label{Theta}
\eea
Due to Bianchi identity $\nabla^\mu G_{\mu\nu}=0$, 
Eq. \rf{eineq} leads to the differential
consequence
\beq
\label{BianchiT}
\nabla^\mu\big[T^{(\phi)}_{\mu\nu}+\coupl \Theta_{\mu\nu}\big]=0.
\eeq
Substituting Eqs. \rf{T} and \rf{Theta} into \rf{BianchiT}, one can check straightforwardly that the differential consequence \rf{BianchiT} is equivalent to \rf{eqmo}.  In other words, Eq. \rf{eqmo} is a differential consequence of Eq. \rf{eineq}.

\section{Isotropic and homogeneous cosmological model}
Let us consider isotropic homogeneous spatially flat cosmological models with the Friedmann–Lema\^{i}tre–Robertson–Walker metric
\begin{equation}
	\label{metric} 
	ds^2=-dt^2+\rma^2(t) d\mathbf{x}^2,
\end{equation}
where $d\mathbf{x}^2$ is the Euclidean metric,
$\rma(t)$ is the scale factor, and $H(t)=\dot{\rma}(t)/\textrm{a}(t)$ is the Hubble parameter. Hereafter, the overdot denotes derivatives with respect to the cosmic time, i.e. $\dot{\rma}(t)=d{\rma}/dt$, etc.
Denoting the present moment of time as $t_0$, we have $\rma_0=\rma(t_0)$ and $H_0=H(t_0)$. 
Supposing homogeneity and isotropy, we also get $\phi=\phi(t)$. 

The general field equations \rf{fieldeq} written for the metric \rf{metric} give the following nontrivial equations:
\bseq\label{genfieldeq}
\bea
\label{00cmpt}
&&3H^2 = 
4\pi{\dot\phi}^2\left(1-9\coupl H^2\right),
\\
\label{11cmpt}
&&\displaystyle
2\dot{H}+3H^2 =
-4\pi\dot{\phi}^2
-4\pi\coupl\dot{\phi}^2\left(2\dot{H}+3H^2 
+4H{\ddot{\phi}}{\dot{\phi}^{-1}}\right),
\\
\label{eqmophi}
&&\frac{1}{\textrm{a}^3}\frac{d}{dt}\left[\textrm{a}^3\dot\phi\left(1- 3\coupl H^2\right)\right]=0,
\eea
\eseq
Here Eq. \rf{00cmpt} is the modified Friedmann equation, i.e. the $tt$-component of \rf{eineq}, Eq. \rf{11cmpt} is the $ii$-component of \rf{eineq} (i=1,2,3), while Eq. \rf{eqmocosm} is the scalar field equation \rf{eqmo}. Note that Eq. \rf{eqmophi} is a differential consequence of Eqs. \rf{00cmpt} and \rf{11cmpt}, therefore only two equations of the system \rf{genfieldeq} are independent.
Integrating Eq. \rf{eqmophi} gives the first integral
\beq\label{eqmocosm}
\dot\phi =\frac{{\cal Q}}{\textrm{a}^3 \left(1-3\coupl H^2\right)},
 \eeq
where ${\cal Q}$ is a constant of integration. In fact, ${\cal Q}$ is the Noether or scalar charge associated with the shift symmetry $\phi\to \phi+\phi_0$.

\subsection{Asymptotic cosmological stages}
As we mentioned in the introduction, cosmological scenarios described by the field equations \rf{genfieldeq} in the theory of gravity with non-minimal derivative coupling have been intensively studied in the literature (see, for example, our works \cite{Sushkov:2009, SarSus:2010, Sushkov:2012, SkuSusTop:2013, MatSus:2015, StaSusVol:2016, StaSusVol:2019, Galeev_etal:2021, Sushkov:2023aya}).
Here let us briefly discuss the key feature of these scenarios. For this purpose, we rewrite the modified Friedmann equation \rf{00cmpt} as follows
\beq \label{modfried}
3H^2 (1+12\pi\coupl\dot\phi^2) =
4\pi{\dot\phi}^2.
\eeq
From the first integral \rf{eqmocosm} one has $\dot\phi\sim \rma^{-3}$, hence $4\pi\coupl\dot\phi^2\gg 1$ in the limit $\rma\to 0$, and the unity in the left-hand side of Eq. \rf{modfried} can be neglected. 
As the result, one has
\beq
H^2= \frac{1}{9\coupl}\,.
\eeq
Therefore, the $\coupl$-term in Eq. \rf{modfried}, dominating at early times, provides the quasi-de Sitter (inflationary) character  of the cosmological evolution such that
\beq
\rma(t)=\rma_i e^{H_\coupl(t-t_i)}, \quad 
\phi(t)=\phi_i e^{-3H_\coupl (t-t_i)}, \quad
\text{where}\quad H_\coupl=\frac{1}{\sqrt{9\coupl}},
\eeq
and $t_i$ is some initial value of time meaning the beginning of inflation.

At late cosmic times, when $\rma(t)$ becomes large, the $\coupl$-terms become small and negligible, so that $4\pi\coupl\dot\phi^2\ll 1$. Now Eq. \rf{modfried} reduces to 
\beq 
3 H^2  = 4\pi{\dot\phi}^2.
\eeq
This is nothing but the standard Friedmann equation describing the late-time evolution of the Universe with the massless scalar field. 
Its solution reads 
\beq
\rma(t)\propto t^{1/3}, \quad H(t)=\frac{1}{3t}, \quad 
\dot\phi\propto \frac{1}{t}.
\eeq

Hereafter, we will call these two asymptotic stages as (i) \textit{quasi-de Sitter (inflationary) stage}, when the non-minimal derivative coupling is dominating, so that $4\pi\coupl\dot\phi^2\gg 1$, and (ii) \textit{post-inflationary stage}, when the non-minimal derivative coupling is negligible, so that $4\pi\coupl\dot\phi^2\ll 1$.

\begin{figure}[htb]
	\centering
	\includegraphics[width=0.5\textwidth]{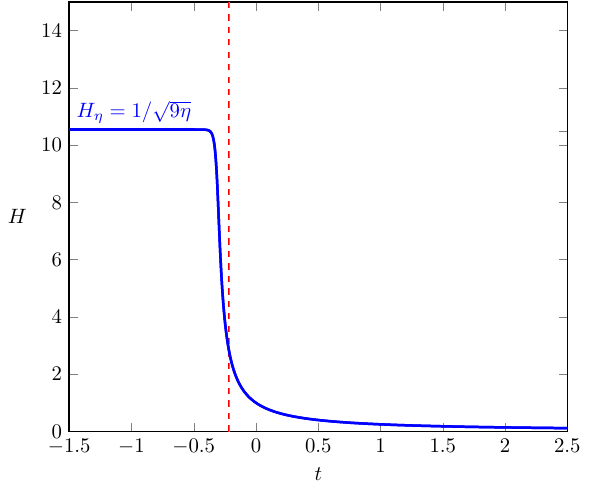}
	\caption{The Hubble parameter $H(t)$ obtained as a numerical solution of the cosmological equations \rf{genfieldeq} with the non-minimal coupling parameter $\eta=0.001$. The vertical dotted line marks the transition moment when $4\pi\coupl\dot\phi^2= 1$.}
	\label{fig:1}
\end{figure}
The entire scenario of the Universe evolution can be obtained numerically as a solution of the cosmological equations \rf{genfieldeq}. The result of numerical analysis is shown in Fig. \ref{fig:1}, where a solution for the Hubble parameter $H(t)$ is presented. The solution clearly reproduces two asymptotic stages: (i) quasi-de-Sitter (inflationary), when $H(t)=H_\eta=1/\sqrt{9\eta}$, and (ii) post-inflationary, when $H(t)=1/3t$. The vertical dotted line in Fig. \ref{fig:1} marks the moment when $4\pi\coupl\dot\phi^2= 1$.   
It is worth noting that the quasi-de Sitter stage is naturally ending and exchanging by the post-inflationary stage as soon as a domination of the non-minimal derivative coupling term comes decreasing during the Universe evolution.
Therefore, the non-minimal derivative coupling provides an inflationary mechanism at early stages of the Universe evolution and naturally describes transitions between various cosmological phases {\em without any fine-tuned potential}.

\subsection{Field equations in conformal time}
A description of cosmological perturbations is more convenient in a conformal time representation of the FLRW metric \rf{metric}:
\begin{equation}
	\label{cmetric} 
	ds^2=\rma^2(\ct)\left[-d\ct^2+d\mathbf{x}^2\right],
\end{equation}
where the conformal time $\ct$ is related with the cosmic time $t$ as $dt=\rma(\ct)d\ct$, $\textrm{a}(\ct)$ is the conformal scale factor, and ${\cal H}(\ct)={\textrm{a}'}(\ct)/\textrm{a}(\ct)$ is the conformal Hubble parameter. Hereafter, the prime denotes derivatives with respect to the  conformal time, i.e. ${\textrm{a}}'(\ct)=d{\textrm{a}}/d\ct$, etc. Note that $H(t)={\cal H}(\ct)/\rma(\ct)$.
Homogeneity and isotropy also suppose that $\phi=\phi(\ct)$. 

Now, the field equations \rf{genfieldeq} reduce to
\bseq\label{confgenfieldeq}
\bea
  \label{conf00cmpt}
  && 3{\cal H}^2= 
  4\pi{\phi}'^2\left(1-9\coupl\rma^{-2} {\cal H}^2\right),
  \\
  \label{conf11cmpt}
  \displaystyle
  && 2{\cal H}'+{\cal H}^2  =  
  -4\pi{\phi}'^2
  \left[1+\coupl\rma^{-2}\left(2{\cal H}'-3{\cal H}^2 
  +4{\cal H} \frac{{\phi}''}{{\phi}'}\right)\right], 
  \\
  \label{confeqmocosm}
	&&\frac{d}{d\ct}\Big[ {\phi}' \rma^2\big(1-3\coupl\rma^{-2} {\cal H}^2\big)\Big] = 0, 
\eea
\eseq
and the first integral \rf{eqmocosm} reduces to
\beq\label{firstint}
{\phi}' \rma^2\big(1-3\coupl \rma^{-2}{\cal H}^2\big)={\cal Q}.
\eeq

In the \textit{quasi-de Sitter (inflationary) stage}, using the relations $dt=\rma(\ct)d\ct$ and $\rma(t)=\rma_i e^{H_\coupl(t-t_i)}$, one can find
\beq
\rma(\ct)=\frac{\rma_i}{1-\rma_i H_\coupl(\ct-\ct_i)},
\eeq
where $\ct_i$ is an arbitrary initial value. Without loss of generality, one can take 
\beq\label{tauin}
\ct_i=-\frac{1}{\rma_i H_\coupl}. 
\eeq
Then, we obtain the following solution:
\beq
\label{dSsol}
\rma(\ct)=-\frac{1}{H_\coupl\ct}, \quad
\calH(\ct)=-\frac{1}{\ct}, \quad 
\phi(\ct)=-\phi_i(\rma_i H_\coupl)^3\,\ct^3.
\eeq
In the \textit{post-inflationary stage} one has $\rma(\ct)\sim \ct^{1/2}$ and ${\cal H(\ct)}=1/2\ct$.

\begin{figure}[htb]
	\centering
	\includegraphics[width=0.5\textwidth]{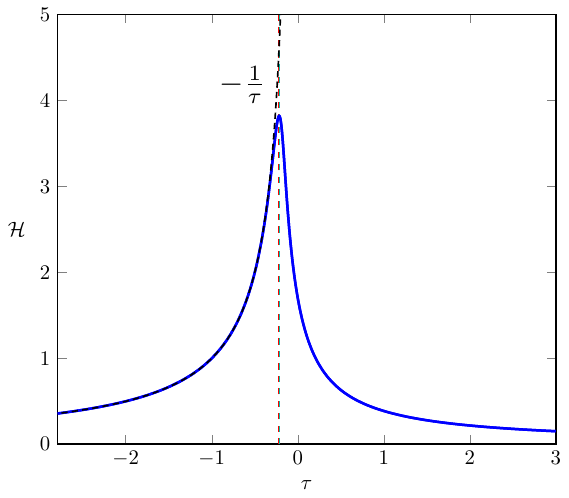}
	\caption{The conformal Hubble parameter $\calH(\ct)$ obtained as a numerical solution of the cosmological equations \rf{confgenfieldeq} with the non-minimal coupling parameter $\eta=0.001$. The vertical dotted line marks the transition moment when $4\pi\coupl\dot\phi^2= 1$. The curved dotted line shows the asymptotic \rf{dSsol}, $\calH(\ct)=-1/\ct$.}
	\label{fig:2}
\end{figure}
In Fig. \ref{fig:2} the conformal Hubble parameter $\calH(\ct)$, obtained as a numerical solution of the equations \rf{confgenfieldeq}, is shown.
The solution clearly reproduces two asymptotic stages: (i) quasi-de-Sitter (inflationary), when $\calH(\ct)=-1/\ct$, and (ii) post-inflationary, when $H(t)=1/2\ct$.

\section{Perturbations}

In the Poisson gauge (see, for example, \cite{Ellis,Ma}) the perturbed FLRW metric \rf{cmetric} is written as follows:
\beq\label{perturbmetric}
ds^2=\rma^2(\ct)\left[-(1+2\Phi) d\ct^2
-2Q_i d\ct dx^i
+\big[ \delta_{ij}(1-2\Psi) + h_{ij} \big]dx^idx^j\right].
\eeq 
Here $\Phi\equiv \Phi(\ct,{\bf x})\ll 1$ and $\Psi\equiv \Psi(\ct,{\bf x})\ll 1$ are scalar perturbations, $Q_i\equiv Q_i(\ct,{\bf x})\ll 1$ are vector perturbations, and $h_{ij}\equiv h_{ij}(\ct,{\bf x})\ll 1$ are tensor perturbations, where $Q_i$ and $h_{ij}$ are transverse, and $h_{ij}$ is a traceless tensor such that
\beq
\nabla^i Q_i=0, \quad \nabla^i h_{ij} = 0,\quad
h^i_{i} = 0.
\eeq
The homogeneous scalar field $\phi(\ct)$ is also perturbed as follows
\beq\label{perturbphi}
\phi(\ct,{\bf x}) = \phi(\ct)(1+\delta\varphi),
\eeq
where $\delta\varphi\equiv \delta\varphi(\ct,{\bf x})\ll 1$ is a perturbation.

Substituting the perturbed metric \rf{perturbmetric} and the scalar field \rf{perturbphi} into the field equations \rf{fieldeq} and taking into account the background relations \rf{confgenfieldeq}, one can obtain the resulting equations for perturbations. Usually, instead of perturbed scalar values $A(\ct,\textbf{x})$, one considers their Fourier modes $A(\ct,\textbf{k})$ defined as follows: 
\beq\label{Fourier}
A(\ct,\textbf{x})=\frac{1}{(2\pi)^{3/2}}\int d\textbf{k}\,e^{i\textbf{k}\textbf{x}} A(\ct,\textbf{k}),
\eeq
where $\textbf{k}$ is a wave vector with components $k_i$ and the length $k$.
It is useful also to use the following substitutions: $\partial_i \to i k_i$, $\Delta \to -{k}^2$.

Further, we separately consider scalar, vector and tensor modes.

\subsection{Scalar modes}
Equations for the scalar modes $\Phi$, $\Psi$, and $\delta\varphi$ read
\begin{eqnarray}
	3\calH(\Psi'+\calH\Phi) +k^2\Psi 
	&=&
	4\pi \left(\phi'^2\Phi-\phi'^2\delta\varphi
	-\phi\phi'\delta\varphi' \right)
	\nonumber\\
	&&
	- 4 \pi \coupl \rma^{-2} \Big[ 
	9\calH \phi'^2\Psi'
	+18\calH^2\phi'^2\Phi  
	+k^2\phi'^2\Psi
	\nonumber\\
	&&
	-9\calH^2\phi\phi'\delta\varphi'
	-9\calH^2\phi'^2\delta\varphi 
	-2k^2\calH\phi\phi'\delta\varphi 
	\Big],
	\label{sc12}
	\\
	\Psi' + \calH \Phi &=& 
	4\pi\phi\phi'\delta\varphi
		-4\pi\coupl \rma^{-2} \Big(\phi'^2\Psi'+3\calH\phi'^2\Phi
	\nonumber\\
	&&  
	-2\calH\phi'^2\delta\varphi
	-2\calH\phi\phi'\delta\varphi'
	+3\calH^2 \phi\phi'\delta\varphi 
	 \Big), 
	\label{sc22}
	\\
	\Psi-\Phi &=& 4 \pi \coupl \rma^{-2} 
	\left[\phi'^2(\Psi+\Phi) + 2\phi\phi''\delta\varphi\right].
	\label{sc32}
\end{eqnarray}
Note that due to isotropy of the background space all perturbation modes depend only on the length $k$ of the wave vector $\textbf{k}$, so that $\Phi\equiv\Phi(\ct,k)$, $\Psi\equiv\Psi(\ct,k)$, $\delta\varphi\equiv\delta\varphi(\ct,k)$. 

The equations \rf{sc12}-\rf{sc32} can be easily analyzed analytically in two asymptotic cases.

\subsubsection{Scalar modes in the post-inflationary stage} 
In this case one can neglect the $\coupl$-terms in Eqs. \rf{sc12}--\rf{sc32}. Then, the equation \rf{sc32} gives that 
\beq
\Phi=\Psi, 
\eeq
and Eqs. \rf{sc12}, \rf{sc22} reduce to
\begin{eqnarray}
	3\calH(\Psi'+\calH\Psi) +k^2\Psi 
	&=&
	4\pi \left(\phi'^2\Psi-\phi'^2\delta\varphi
	-\phi\phi'\delta\varphi' \right),
	\label{postinf1}
	\\
	\Psi' + \calH \Psi &=& 
	4\pi\phi\phi'\delta\varphi.
	\label{postinf2}
\end{eqnarray}
Using \rf{postinf2} to exclude $\delta\varphi$ from Eq.\rf{postinf1} and taking into account that $\calH=1/2\ct$ and $3\calH^2=4\pi\phi'^2$, we obtain the following equation for the scalar perturbation mode $\Psi(\ct,{k})$:
\begin{equation}\label{sm_postinf}
	\Psi''+\frac{3}{\ct}\Psi'+k^2\Psi=0.
\end{equation}
Its general solution is as follows:
\beq\label{sm_exactsol}
\Psi(\ct,{k})=\frac{C^{(s)}_{1,k}}{k\ct}J_1(k\ct)+\frac{C^{(s)}_{2,k}}{k\ct}Y_1(k\ct),
\eeq
where $J_1(k\ct)$ and $Y_1(k\ct)$ are the Bessel functions, and $C^{(s)}_{1,k}$ and $C^{(s)}_{2,k}$ are constants of integration. Note that $C^{(s)}_{1,k}$,  $C^{(s)}_{2,k}$ are not arbitrary. Their values are determined by a previous stage of evolution of the modes. 

For \textit{long-wavelength scalar modes},\footnote{Taking into account that $\calH=1/2\ct$ in the post-inflationary stage, one can recast the condition $k\ct\ll1$ as $k^{-1}\gg\calH^{-1}$, so the long-wavelength modes are \textit{superhorizon} ones. Analogously, the short-wavelength modes, $k\ct\gg1$, are \textit{subhorizon} ones.} when $k\ct\ll 1$, the solution  \rf{sm_exactsol} reduces to
\beq\label{lwsmdecay}
\Psi(\ct,{k})\approx \frac12{C^{(s)}_{1,k}}-\frac{2C^{(s)}_{2,k}}{\pi(k\ct)^2},
\eeq
and for \textit{short-wavelength scalar modes}, when $k\ct\gg 1$, one has
\beq\label{swsmdecay}
\Psi(\ct,{k})\propto \frac{1}{(k\ct)^{3/2}}\cos(k\ct+\alpha^{(s)}_k),
\eeq
where $\alpha^{(s)}_k$ is an arbitrary phase. 

Note that the result obtained for scalar perturbation modes in the post-inflationary stage is expectable and well-known for cosmology with an ordinary massless scalar field (see, for example, \cite{Padmanabhan:2010}). A completely novel result is expected for the modes in the early quasi-de Sitter stage.  

\subsubsection{Scalar modes in the quasi-de Sitter (inflationary) stage}
First, let us rewrite Eq. \rf{sc32} as follows
\beq
(1-4\pi\coupl\rma^{-2}\phi'^2)\Psi-(1+4\pi\coupl\rma^{-2}\phi'^2)\Phi=
8\pi\coupl\rma^{-2}\phi\phi''\delta\varphi.
\eeq
Taking into account that in the quasi-de Sitter (inflationary) stage $4\pi\coupl\rma^{-2}\phi'^2\gg1$ and using the solution \rf{dSsol} for $\phi(\ct)$, one can obtain from the above equation that
\beq\label{varphi}
\delta\varphi=- \frac34 (\Psi+\Phi).
\eeq 
Performing analogous manipulations with Eqs. \rf{sc12}, \rf{sc22}, substituting the background expressions for $\rma(\ct)$, $\calH(\ct)$, $\phi(\ct)$ given by Eq. \rf{dSsol}, and taking into account the relation \rf{varphi}, we can obtain the following equations describing the scalar perturbations in the quasi-de Sitter stage:
\begin{eqnarray}
\ct\Psi'-\Phi-\frac{k^2\ct^2}{18}(\Psi-\Phi)&=&0,
\\
\ct\Psi'-\ct\Phi'-6\Phi&=&0.
\end{eqnarray} 
A general solution of these equations yields
\begin{eqnarray}
	\Psi(\ct,{k})&=&
	\frac{D^{(s)}_{1,k}}{\xi^5}(\xi^3-9\xi+9\sqrt{3})e^{\xi/\sqrt{3}}	
	\nonumber\\	
	&&+\frac{D^{(s)}_{2,k}}{\xi^5}(\xi^3-9\xi-9\sqrt{3})e^{-\xi/\sqrt{3}},
	\label{solPsi}\\
	\Phi(\ct,{k})&=&\frac{D^{(s)}_{1,k}}{\xi^5}(\xi^3-6\sqrt{3}\xi^2+45\xi-45\sqrt{3})e^{\xi/\sqrt{3}}
	\nonumber\\
	&&+\frac{D^{(s)}_{2,k}}{\xi^5}(\xi^3+6\sqrt{3}\xi^2+45\xi+45\sqrt{3})e^{-\xi/\sqrt{3}},
	\label{solPhi}
\end{eqnarray}
where the dimensionless value $\xi=-k\ct$ was introduced, and $D^{(s)}_{1,k}$ and $D^{(s)}_{2,k}$ are constants of integration. Here, it is necessary to emphasize that values of $D^{(s)}_{1,k}$ and $D^{(s)}_{2,k}$ have to choose so that initial scalar perturbations are small, i.e. $\Psi_{i,k}\equiv\Psi(\ct_i,k)\ll 1$ and $\Phi_{i,k}\equiv\Phi(\ct_i,k)\ll 1$, where $\ct_i=-{1}/{\rma_i H_\coupl}$ is the initial moment of time (see Eq. \rf{tauin}). In practice, performing numerical calculations, we fix initial values of $\Psi_{i,k}$, $\Phi_{i,k}$, and determine $D^{(s)}_{1,k}$ and $D^{(s)}_{2,k}$ from Eqs. \rf{solPsi}, \rf{solPhi}, assuming   $\xi=\xi_i=-k\ct_i$. 
Note also that, taking into account that $\calH=-1/\ct$ in the quasi-de Sitter (inflationary) stage, one gets $\xi=k\calH^{-1}$. Then, modes with $\xi\ll1$ correspond to superhorizon ones such that $k^{-1}\gg\calH^{-1}$, and modes with $\xi\gg1$ correspond to subhorizon ones such that $k^{-1}\ll\calH^{-1}$. 

Now, let us discuss an asymptotic behavior of scalar modes \rf{solPsi} and \rf{solPhi}. As was mentioned above, initially, at $\ct=\ct_i=-{1}/{\rma_i H_\coupl}$, the scalar perturbation modes are small, $\Psi_{i,k}\ll 1$ and $\Phi_{i,k}\ll 1$. During the inflationary stage $\tau$ is increasing from $\tau_i$ to $\tau_e$, where $\ct_e$ is a time when the inflationary stage ends. Asymptotic values    $\Psi_{e,k}\equiv\Psi(\ct_e,k)$ and $\Phi_{e,k}\equiv\Phi(\ct_e,k)$ at the end of the inflationary stage are different for superhorizon and subhorizon modes.
%
Actually, let us consider the long-wavelength modes such that at the end of the inflationary stage one has $\xi_e\ll 1$. 
%
In the limit 
$\xi\to0$, the scalar modes \rf{solPsi} and \rf{solPhi} behave as
\bea
\Psi(\ct,k) &\approx& 
-\frac{9\sqrt{3}}{\xi^5}(D^{(s)}_{1,k}-D^{(s)}_{2,k}),
\\
\Phi(\ct,k) &\approx& \frac{45\sqrt{3}}{\xi^5}(D^{(s)}_{1,k}-D^{(s)}_{2,k}).
\eea  
It is seen that in the very specific case, when $D^{(s)}_{1,k}=D^{(s)}_{2,k}$, the modes $\Psi(\ct,k)$ and $\Phi(\ct,k)$ remain finite in the limit $\xi\to 0$. However, generally the modes are increasing 
and become as large as $\Psi_{e,k},\Phi_{e,k}
\sim (D^{(s)}_{1,k}-D^{(s)}_{2,k})\xi_{e}^{-5}$. 
Therefore, the scalar modes are amplified during the quasi-de Sitter (inflationary) stage. A degree of amplification could be estimated as follows. Supposing that the value $\xi_i$ is also small, i.e. $\xi_i\ll1$, we get from Eqs. \rf{solPsi}, \rf{solPhi} for the initial perturbations $\Psi_{i,k},\Phi_{i,k}$ the following estimate:  
$\sim (D^{(s)}_{1,k}-D^{(s)}_{2,k})\xi_{i}^{-5}$. 
Therefore,
\beq \label{As}
\frac{\Psi_{e,k}}{\Psi_{i,k}} \approx \frac{\Phi_{e,k}}{\Phi_{i,k}}
\approx \,\left(\frac{\xi_i}{\xi_e}\right)^5=
\,\left(\frac{\ct_i}{\ct_e}\right)^5\equiv {\cal A}^{(s)}. 
\eeq
where ${\cal A}^{(s)}=(\ct_i/\ct_e)^5$ is the amplifying factor. It is necessary to stress that the amplifying factor does not depend on $k$ and is the same for all long-wavelength modes providing the condition $\xi\ll1$, or $k^{-1}\gg\calH^{-1}$. For given the amplifying factor ${\cal A}$, one can always choose appropriately small initial perturbations $\Psi_{i,k}\ll1$ and $\Phi_{i,k}\ll1$ so that final perturbations $\Psi_{e,k}={\cal A}^{(s)} \Psi_{i,k}$ and $\Phi_{e,k}={\cal A}^{(s)} \Phi_{i,k}$ remain also small. This guarantees that the linear perturbation method is adequately applicable for an analysis of evolution of the long-wavelength modes during the quasi-de Sitter (inflationary) stage. 

Now consider the short-wavelength modes, $\xi\gg 1$. In the limit $\xi\to\infty$, the asymptotic of scalar modes given by Eqs. \rf{solPsi} and \rf{solPhi} is 
\beq
\Psi(\ct,k) \approx \Phi(\ct,k) \approx \frac{D_{1,k}^{(s)}}{\xi^2}e^{\xi/\sqrt{3}}, 
\eeq
so at the end of the inflationary stage one has 
$
\Psi_{e,k} \approx \Phi_{e,k} \approx {D_{1,k}^{(s)}}{\xi_e^{-2}}e^{\xi_e/\sqrt{3}}. 
$
Note that the short-wavelength modes remain being subhorizon during the whole inflationary stage, so that $\xi_i\gg1$, and, analogously, one has
$
\Psi_{i,k} \approx \Phi_{i,k} \approx {D_{1,k}^{(s)}}{\xi_i^{-2}}e^{\xi_i/\sqrt{3}}. 
$
Therefore, 
\beq
\frac{\Psi_{e,k}}{\Psi_{i,k}} \approx \frac{\Phi_{e,k}}{\Phi_{i,k}}
\approx \,\left(\frac{\xi_i}{\xi_e}\right)^2 e^{(\xi_e-\xi_i)/\sqrt{3}}=
\,\left(\frac{\ct_i}{\ct_e}\right)^2 e^{k(\ct_e-\ct_i)/\sqrt{3}}\equiv {\cal B}_k^{(s)}, 
\eeq
where the amplifying factor ${\cal B}_k^{(s)}=({\ct_i}/{\ct_e})^2 e^{k(\ct_e-\ct_i)/\sqrt{3}}\,$ is now exponentially depending on the wave number $k$. This means that there exists some critical value $k_*^{(s)}$ such that all modes with $k\geq k_*^{(s)}$ grow so intensively that
\bea
\Psi_{e,k\geq k_*^{(s)}} &=& {\cal B}_{k\geq k_*^{(s)}}^{(s)}\Psi_{i,k\geq k_*^{(s)}}\geq 1,
\nonumber\\
\Phi_{e,k\geq k_*^{(s)}} &=& {\cal B}_{k\geq k_*^{(s)}}^{(s)}\Phi_{i,k\geq k_*^{(s)}}\geq 1.
\eea 
That is the linear perturbation analysis is not applicable for studying an evolution of the short-length modes with $k\geq k_*$ (at least in the whole inflationary stage). Instead, one has to use some nonlinear methods. It should be noted that, from a physical point of view, the transition to nonlinear evolution of scalar perturbation modes means the appearance of large-scale structures in the Universe with scales of $l\leq l_*=k_*^{-1}$.

%
\begin{figure}[htb]
	\centering
	\includegraphics[width=0.45\textwidth]{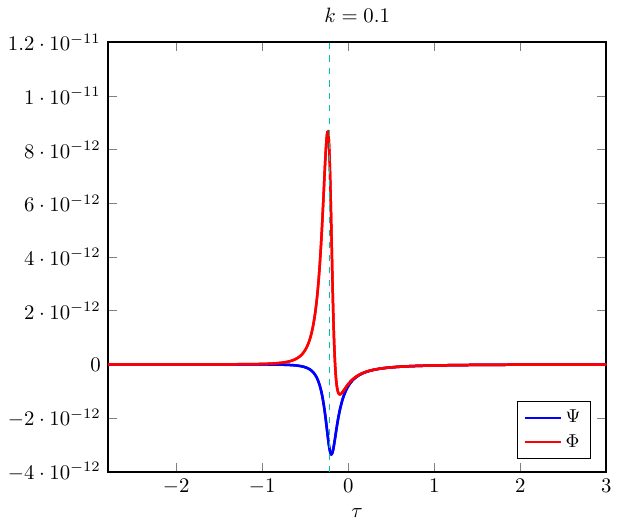}
	\includegraphics[width=0.45\textwidth]{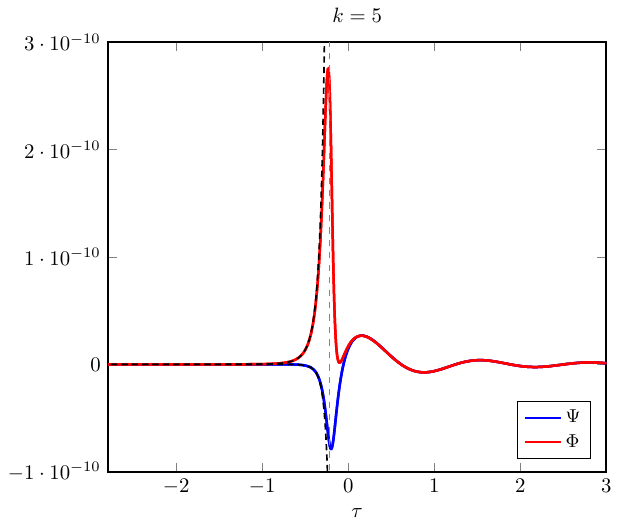}
	\\
	\includegraphics[width=0.45\textwidth]{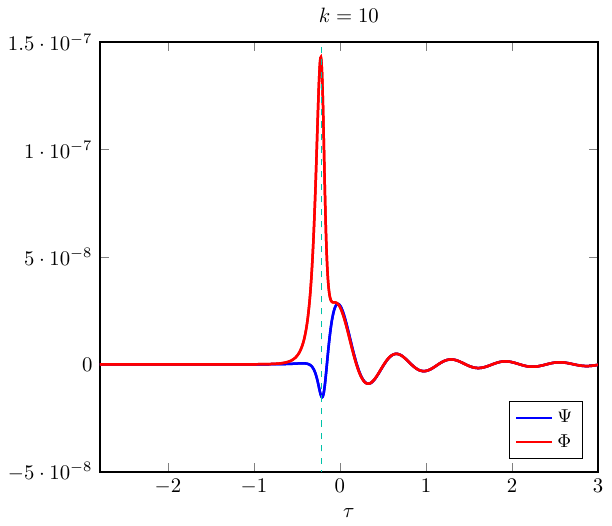}
	\includegraphics[width=0.45\textwidth]{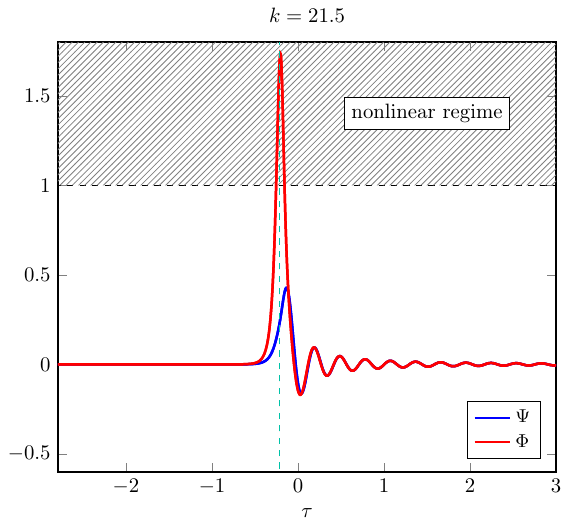}
	\caption{Graphs for the scalar modes $\Psi(\ct,k)$ (blue line) and $\Phi(\ct,k)$ (red line) are given for $k=0.1, 5, 10, 21.5$. Initial values for numerical analysis are the following: $\ct_i\approx-2.79112$, $\Psi_i\equiv \Psi(\ct_i,k)=10^{-16}$, $\Phi_i\equiv \Phi(\ct_i,k)=10^{-16}$. The coupling parameter is $\coupl=0.001$. The vertical dotted line labels the moment $\ct_e\approx-0.25525$, when the quasi-de Sitter (inflationary) stage ends and the post-inflationary one begins. The moment $\ct_e$ is determined from the condition $4\pi\coupl\rma^{-2}\phi'^2=1$.}
	\label{fig:smodes}
\end{figure}
\subsubsection{General behavior of the scalar modes \label{numproc}}
Here we present results of numerical analysis of the system \rf{sc12}-\rf{sc32}, which describes the entire evolution of the scalar modes. The algorithm of numerical analysis is the following. First we numerically integrate the unperturbed cosmological equations \rf{confgenfieldeq} forward and backward in the conformal time $\ct$, using the initial conditions $\rma_0\equiv\rma(\ct_0)=1$, $\calH_0\equiv\calH(\ct_0)=1$, $\ct_0=0$. Then, we shift an initial moment of time $\ct_i$ so that the Hubble function $\calH(\ct)$ obtained numerically would coincide with the asymptotic solution $\calH(\ct)=-1/\ct$ in the quasi-de Sitter (inflationary) stage (see Fig. \ref{fig:2}). Next, we numerically integrate the system \rf{sc12}-\rf{sc32}, using an initial conditions $\Psi(\ct_i,k)=\Psi_{i,k}$, $\Phi(\ct_i,k)=\Phi_{i,k}$. An initial value $\delta\varphi_i$ is found from the algebraic constraint \rf{sc32}. 

Everywhere in this paper we demonstrate numerical results obtained for the coupling parameter $\coupl=0.001$ and the initial values $\Psi_i=10^{-16}$, $\Phi_i=10^{-16}$ given for the initial moment of the conformal time $\ct_i\approx-2.79112$. 
Results are presented graphically in Fig. \ref{fig:smodes}. In plots a vertical dotted line labels to the moment $\ct_e\approx-0.25525$, when the quasi-de Sitter (inflationary) stage ends and the post-inflationary one begins. The moment $\ct_e$ is determined from the condition $4\pi\coupl\rma^{-2}\phi'^2=1$. Correspondingly, in the region $\ct<\ct_e$, where $4\pi\coupl\rma^{-2}\phi'^2\gg1$, one has the quasi-de  Sitter (inflationary) stage of the Universe evolution, and in the region $\ct>\ct_e$, where $4\pi\coupl\rma^{-2}\phi'^2\ll1$, the post-inflationary stage. It should be emphasized that the numerical results perfectly confirm the asymptotic analysis and exactly reproduce asymptotic solutions \rf{sm_exactsol}, \rf{solPsi} and \rf{solPhi}, and hence all asymptotic properties established above. In particular, it has been shown that the long-wavelength scalar modes are amplifying by the factor ${\cal A}^{(s)}=(\ct_i/\ct_e)^5$ during the inflationary stage. Given  $\ct_i\approx-2.79112$ and $\ct_e\approx-0.25525$, one has ${\cal A}^{(s)}\approx 1.56\times 10^5$. The first panel in Fig. \ref{fig:smodes} given for $k=0.1$ illustrates this property. For short-wavelength modes it has been shown that they are amplifying by the factor  ${\cal B}_k^{(s)}=({\ct_i}/{\ct_e})^2 e^{k(\ct_e-\ct_i)/\sqrt{3}}$, and hence there exists a critical value $k_*^{(s)}$ such that all perturbation modes with $k\geq k_*^{(s)}$ go to non-linear regime. Numerically it was found that  for scalar modes $k_*^{(s)}\approx21.48$, and the forth panel in Fig. \ref{fig:smodes} illustrates this result.

\subsection{Tensor modes}
As usually, an arbitrary transversal traceless tensor $h_{ij}$ can be
represented as a linear combination of two basic tensors $e^{(+)}_{ij}$ and $e^{(\times)}_{ij}$ with helicities $+2$ and $-2$, respectively, so that
\beq
h_{ij}=\sum_{A=+,\times}e^{(A)}_{ij} h^{(A)},
\eeq
where $h^{(A)}$ are amplitudes.
Applying the Fourier transformation \rf{Fourier} for $h^{(A)}$, one can obtain the following equation for tensor perturbation modes $h^{(A)}(\ct,k)$: 
\beq
(1+4\pi\coupl\rma^{-2}\phi'^2) h''
+2\left(\calH+4\pi\coupl\rma^{-2}\phi'\phi''\right)h'
+{k^2}(1-4\pi \coupl\rma^{-2}\phi'^2) h=0,
\label{amplitude}
\eeq
where for simplicity we have omitted the index $A$. 


\subsubsection{Tensor modes in the post-inflationary stage} 
Neglecting the $\coupl$-terms in Eq. \rf{amplitude} and substituting $\calH=1/2\ct$, we obtain the following equation for tensor modes in the post-inflationary stage: 
\beq
h'' +\frac{1}{\ct} h'+{k^2} h=0.
\eeq
Its general solution is
\beq\label{tm_exactsol}
h(\ct,{k})={C^{(t)}_{1,k}}J_0(k\ct)+{C^{(t)}_{2,k}}Y_0(k\ct),
\eeq
where $J_0(k\ct)$ and $Y_0(k\ct)$ are the Bessel functions, and $C^{(t)}_{1,k}$ and $C^{(t)}_{2,k}$ are constants of integration.

For \textit{long-wavelength tensor modes}, when $k\ct\ll 1$, the solution  \rf{tm_exactsol} reduces to
\beq
h(\ct,{k})\approx C^{(t)}_{1,k}+\frac{2}{\pi}C^{(t)}_{2,k}\left(\ln\frac{k\ct}{2}+\gamma\right),
\eeq
where $\gamma$ is Euler's constant, and for \textit{short-wavelength tensor modes}, when $k\ct\gg 1$, one has
\beq
h(\ct,{k})\propto \frac{1}{(k\ct)^{1/2}}\cos(k\ct+\alpha^{(t)}_k),
\eeq
where $\alpha^{(t)}_k$ is an arbitrary phase. 

\subsubsection{Tensor modes in the quasi-de Sitter (inflationary) stage}
Taking into account that $4\pi\coupl\rma^{-2}\phi'^2\gg1$ and substituting into Eq. \rf{amplitude} the background expressions for $\rma(\ct)$, $\calH(\ct)$, $\phi(\ct)$ given by Eq. \rf{dSsol}, we obtain the following equation describing the tensor perturbations in the quasi-de Sitter (inflationary) stage:
\beq
h''+\frac{4}{\ct}h'-k^2h=0.
\eeq
Its general solution reads
\beq\label{solh}
h(\ct,k)=\frac{D^{(t)}_{1,k}}{\xi^3}(\xi-1)e^{\xi}
+\frac{D^{(t)}_{2,k}}{\xi^3}(\xi+1)e^{-\xi},
\eeq
where $\xi=-k\ct$. The constants of integration $D^{(t)}_{1,k}$ and $D^{(t)}_{2,k}$ are chosen so that initial tensor modes are small, $h_{i,k}\equiv h(\ct_i,k)\ll 1$, where $\ct_i=-1/\rma_i H_\eta$.

Analogously with scalar modes, long- and short-wavelength tensor modes behave differently at the end of the quasi-de Sitter (inflationary) stage. For the long-wavelength modes with $\xi\ll1$ one has in the limit $\xi\to0$ 
%
\beq\label{as_t_long}
h(\ct,k)\approx 
-\frac{1}{\xi^3}(D^{(t)}_{1,k}-D^{(t)}_{2,k}).
\eeq
Therefore, the long-wavelength tensor modes are amplified during the inflationary stage, and the degree of amplification is the following
\beq \label{At}
\frac{h_{e,k}}{h_{i,k}} \approx \left(\frac{\xi_i}{\xi_e}\right)^3=
\,\left(\frac{\ct_i}{\ct_e}\right)^3\equiv {\cal A}^{(t)}. 
\eeq
where ${\cal A}^{(t)}=(\ct_i/\ct_e)^3$ is the amplifying factor.

For the short-wavelength modes with $\xi\gg 1$, one has in the limit $\xi\to\infty$ 
\beq\label{as_t_short}
h(\ct,k) \approx \frac{D_{1,k}^{(t)}}{\xi^2}e^{\xi} 
\eeq
Therefore, the short-wavelength tensor modes are amplified during the inflationary stage, and the degree of amplification is the following
\beq
\frac{h_{e,k}}{h_{i,k}} \approx \,
\left(\frac{\xi_i}{\xi_e}\right)^2 e^{(\xi_e-\xi_i)}=
\,\left(\frac{\ct_i}{\ct_e}\right)^2 e^{k(\ct_e-\ct_i)}\equiv {\cal B}_k^{(t)}, 
\eeq
where the amplifying factor ${\cal B}_k^{(t)}=({\ct_i}/{\ct_e})^2 e^{k(\ct_e-\ct_i)}\,$ is now exponentially depending on the wave number $k$. This means that there exists some critical value $k_*^{(t)}$ such that all modes with $k\geq k_*^{(t)}$ go to non-linear regime:
\beq
h_{e,k\geq k_*^{(t)}} = {\cal B}_{k\geq k_*^{(t)}}^{(t)} h_{i,k\geq k_*^{(t)}}\geq 1.
\eeq 
Note that, comparing ${\cal B}_k^{(s)}$ and ${\cal B}_k^{(t)}$, one can conclude that $k_*^{(s)}=\sqrt{3}\,k_*^{(t)}$. This means that a non-linear regime begins for  scalar modes which are shorter by the factor $\sqrt{3}$ than analogous tensor modes.

\subsubsection{General behavior of the tensor modes}
\begin{figure}[htb]
	\centering
	\includegraphics[width=0.45\textwidth]{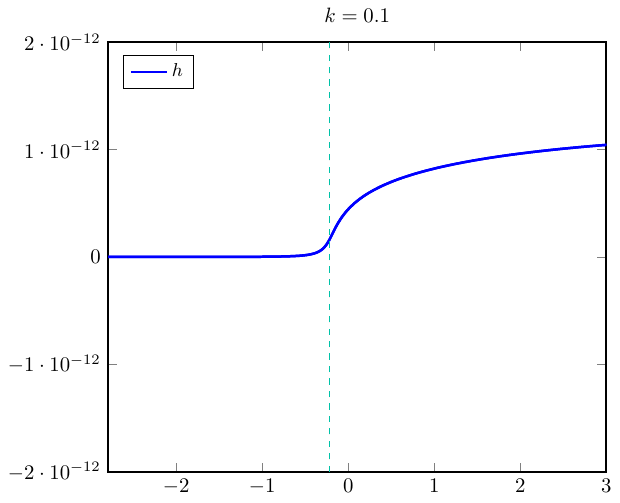}
	\includegraphics[width=0.45\textwidth]{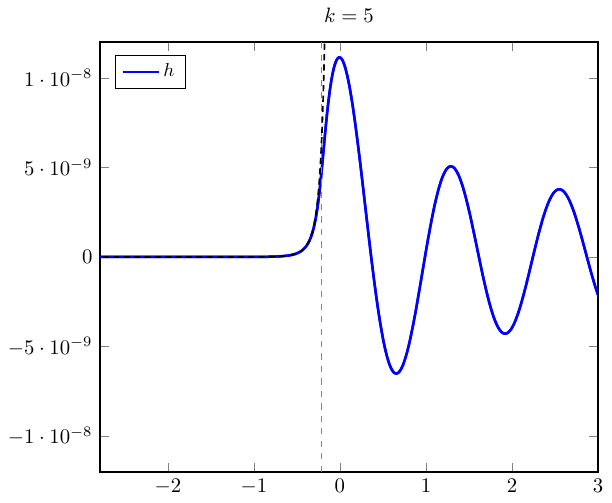}
	\\
	\includegraphics[width=0.45\textwidth]{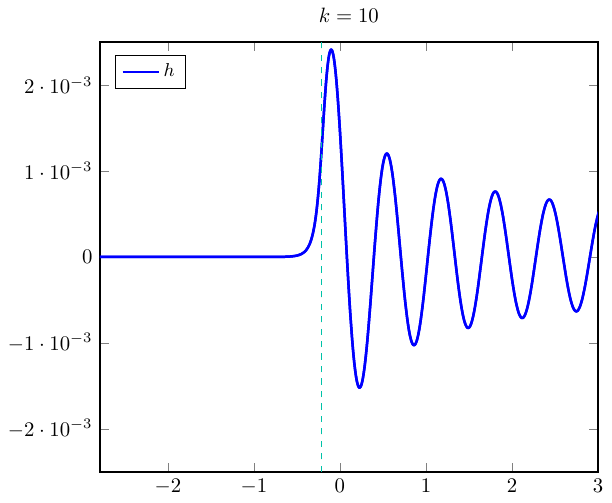}
	\includegraphics[width=0.45\textwidth]{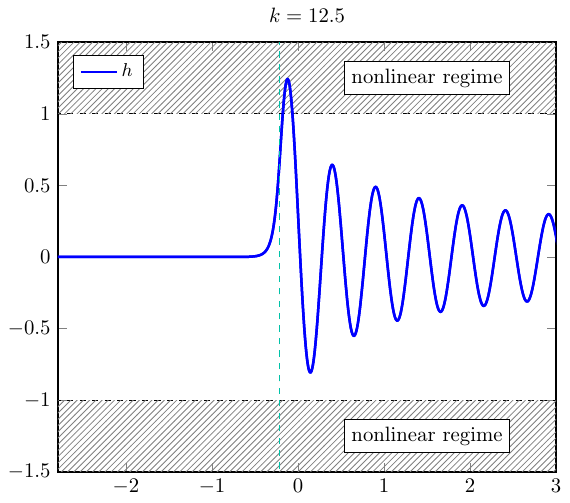}
	\caption{Graphs for the tensor modes $h(\ct,k)$ are given for  $k=0.1, 5, 10, 12.5$. Initial values for numerical analysis are the following: $\ct_i\approx-2.79112$, $h_i\equiv h(\ct_i,k)=10^{-16}$, $h'_i\equiv h'(\ct_i,k)=10^{-16}$. The coupling parameter is $\coupl=0.001$. The vertical dotted line labels the moment $\ct_e\approx-0.25525$, when the quasi-de Sitter (inflationary) stage ends and the post-inflationary one begins. The moment $\ct_e$ is determined from the condition $4\pi\coupl\rma^{-2}\phi'^2=1$.}
	\label{fig:tmodes}
\end{figure}

Here we present results of numerical analysis of the equation \rf{amplitude}, which describes the entire evolution of the tensor modes. The numerical procedure is described in detail in the corresponding subsection for scalar modes.
Results are presented graphically in Fig. \ref{fig:tmodes}.
Note that the numerical results perfectly confirm the asymptotic analysis and exactly reproduce asymptotic solutions \rf{as_t_long} and \rf{as_t_short}, and hence all asymptotic properties established above. In particular, it has been shown that the long-wavelength tensor modes are amplifying by the factor ${\cal A}^{(t)}=(\ct_i/\ct_e)^3$ during the inflationary stage. Given  $\ct_i\approx-2.79112$ and $\ct_e\approx-0.25525$, one has ${\cal A}^{(t)}\approx 1.3\times 10^3$. The first panel in Fig. \ref{fig:tmodes} given for $k=0.1$ illustrates this property. For short-wavelength modes it has been shown that they are amplifying by the factor  ${\cal B}_k^{(t)}=({\ct_i}/{\ct_e})^2 e^{k(\ct_e-\ct_i)}$, and hence there exists a critical value $k_*^{(t)}$ such that all perturbation modes with $k\geq k_*^{(t)}$ go to non-linear regime. Numerically it was found that for tensor modes $k_*^{(t)}\approx 12.4$, and the forth panel in Fig. \ref{fig:tmodes} illustrates this result.

\subsection{Vector modes}
Let us represent a vector mode as $Q_i(\ct,k)=Q(\ct,k) S_i$, where $Q(\ct,k)$ is a scalar function, and $S_i$ is a constant vector such that $\Delta S_i=k^2 S_i$. Then, the equation for vector modes $Q(\ct,k)$ takes the following form:
\beq\label{vm}
(1+4\pi\coupl\rma^{-2}\phi'^2)Q' +2(\calH+4\pi\coupl\rma^{-2}\phi'\phi'')Q=0.
\eeq

\subsubsection{Vector modes in the post-inflationary stage} 
Neglecting the $\coupl$-terms in Eq. \rf{vm} and substituting $\calH=1/2\ct$, we obtain the following equation for vector modes in the post-inflationary stage: 
\beq
Q' +\frac{1}{\ct} Q=0.
\eeq
From here we get
\beq\label{vmdecay}
Q=\frac{C^{(v)}_k}{k\ct}\propto \frac{1}{\rma^2},
\eeq
where $C^{(v)}_k$ is a constant of integration. Therefore, as was expected, the vector modes decay as $\rma^{-2}$ in the post-inflationary stage. 

Note that in the framework of GR this fact serves as an argument to ignore completely vector perturbations. Actually, the primordial vector perturbations may have significant amplitudes at present only if they were originally very large. However, in GR there is no reason to expect such large primordial vector perturbations. The situation is changed crucially if one takes into account the primary quasi-de Sitter (inflationary) stage, which is realized in the theory of gravity with non-minimal derivative coupling.    

\subsubsection{Vector modes in the quasi-de Sitter (inflationary) stage}
Taking into account that $4\pi\coupl\rma^{-2}\phi'^2\gg1$ and substituting into Eq. \rf{amplitude} the background expressions for $\rma(\ct)$, $\calH(\ct)$, $\phi(\ct)$ given by Eq. \rf{dSsol}, we obtain the following equation describing the vector perturbations in the quasi-de Sitter (inflationary) stage:
\beq
Q'+\frac{4}{\ct}Q=0.
\eeq
Its solution is
\beq\label{solq}
Q(\ct,k)=\frac{D^{(v)}_{k}}{\xi^4}=\frac{D^{(v)}_{k}}{(k\ct)^4},
\eeq
where $\xi=-k\ct$, and $D^{(v)}_{k}$ is a constant of integration, which is chosen so that initial vector modes are small, $Q(\ct_i,k)\ll 1$, where $\ct_i=-1/\rma_i H_\eta$. Taking into account that $\rma(\ct)=-1/H_\eta \ct$, one has $Q(\ct,k)\propto \rma^4$. That is the vector modes are increasing as $\rma^4$ at the inflationary stage, in distinct with the post-inflationary one, where the modes decaying as $\rma^{-2}$. 
During the inflationary stage 
$\ct$ is increasing from $\ct_i$ to $\ct_e$, where $\ct_e$ is a time when the inflationary stage ends and the post-inflationary one begins. Correspondingly,
the vector modes are increasing and become as large as $Q(\ct_e,k)= D^{(v)}_{k}(k\ct_{e})^{-4}$. Using the relation $D^{(v)}_{k}=(k\ct_i)^4 Q(\ct_i,k)$, we obtain the following estimate:
\beq\label{Av}
\frac{Q(\ct_e,k)}{Q(\ct_i,k)}=\left(\frac{\ct_i}{\ct_e}\right)^4 
\equiv {\cal A}^{(v)}.
\eeq
Therefore, during the inflationary stage \textit{all} vector modes, \textit{regardless} of $k$, are amplified by a factor of ${\cal A}^{(v)}=\left({\ct_i}/{\ct_e}\right)^4$. 

\subsubsection{General behavior of the vector modes}
\begin{figure}[htb]
	\centering
	\includegraphics[width=0.5\textwidth]{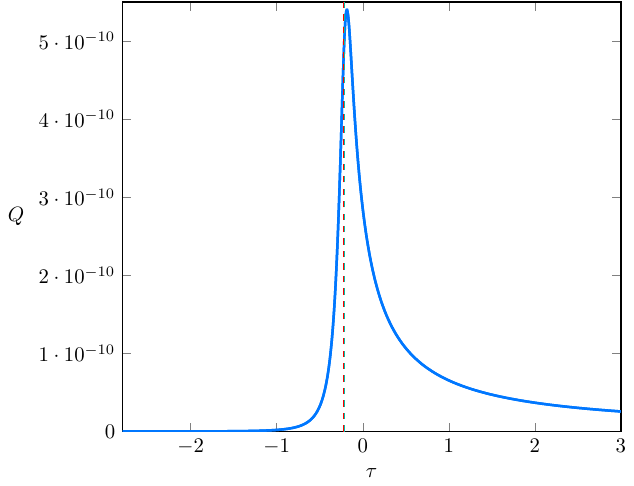}
	\caption{Graphs for the vector modes $Q(\ct,k)$ are given for three values of $k=0.1, 1, 10$. Initial values for numerical analysis are the following: $\ct_i\approx-2.79112$, $Q_i\equiv Q(\ct_i,k)=10^{-16}$. The coupling parameter is $\coupl=0.001$. The vertical dotted line labels the moment $\ct_e\approx-0.25525$, when the quasi-de Sitter (inflationary) stage ends and the post-inflationary one begins. The moment $\ct_e$ is determined from the condition $4\pi\coupl\rma^{-2}\phi'^2=1$.}
	\label{fig:vmodes}
\end{figure}

In this section we present results of numerical analysis of the equation \rf{vm}, which describes the entire evolution of the vector modes. Results are presented graphically in Fig. \ref{fig:vmodes}.
The numerical procedure is described in detail in the corresponding subsection for scalar modes.
Results are presented graphically in Fig. \ref{fig:tmodes}.
Note that the numerical results perfectly confirm the asymptotic analysis and exactly reproduce asymptotic solutions \rf{vmdecay} and \rf{solq}, and hence all asymptotic properties established above. In particular, it has been shown that all vector modes are amplifying by the factor ${\cal A}^{(v)}=(\ct_i/\ct_e)^4$ during the inflationary stage. Given  $\ct_i\approx-2.79112$ and $\ct_e\approx-0.25525$, one has ${\cal A}^{(v)}\approx 1.43\times 10^4$.
%

\section{Discussion and concluding remarks}
In the paper we have explored perturbations in the isotropic and homogeneous cosmological model with the spatially flat Friedmann–Lema\^{i}tre–Robertson–Walker metric in the theory of gravity with non-minimal derivative coupling (see the action \rf{action}). 
For this aim we have derived a complete set of equations which describe an evolution of scalar, vector and tensor modes of perturbations. All modes was analyzed analytically in two asymptotic stages of the Universe evolution: (i) quasi-de Sitter (inflationary), and (ii) post-inflationary. 
Note that the quasi-de Sitter (inflationary) stage is realized on early times, when the non-minimal derivative coupling interaction is dominating. While, the post-inflationary stage is realized on late times, when the influence of non-minimal derivative coupling on the Universe evolution completely disappears, and this naturally leads to the transition to the standard cosmological evolution.

For all perturbation modes (scalar, tensor, vector) in the post-inflationary stage we have reproduced results which are expectable and well-known for cosmology with an ordinary massless scalar field (see Eqs. \rf{sm_exactsol}, \rf{tm_exactsol}, \rf{vmdecay}). At the same time, a behavior of all modes in the quasi-de Sitter (inflationary) stage, which is driven by the non-minimal derivative coupling interaction, is completely novel and distinct from that in the standard `potential' inflation. 

We found that perturbation modes, including vector ones, are amplified in the quasi-de Sitter (inflationary) stage. 
A degree of amplification is different for different types of modes (scalar, tensor, or vector), and depends also on a mode wavelength.
The degree of amplification of the \textit{long-wavelength} ($k\ct\ll1$) scalar and tensor modes, and also vector modes is approximately determined by the factor ${\cal A}^{(a)}=(\ct_i/\ct_e)^n$, where the label $n=3$ for tensor ($a=t$), $n=4$ for vector ($a=v$), and $n=5$ for scalar ($a=s$) modes (see Eqs. \rf{At},\rf{Av}, and \rf{As}). Here, $\ct_i$ is a time of beginning of the inflationary stage, and $\ct_e$ is a time when the inflationary stage ends and the post-inflationary one begins. Let us recall that $\ct$ is the conformal time, $\ct_i=-1/\rma_i H_\eta$, and during the inflationary stage $\ct$ is increasing from $\ct_i$ to $\ct_e$. The time $\ct_e$, when the inflationary stage ends, depends on parameters of the model and is obtained numerically. Our numerical analysis\footnote{Note that in this work we presents numerical results obtained for the coupling parameter $\coupl=0.001$. For other values of $\coupl$ results are qualitatively the same.} 
gave $\ct_i\approx-2.79112$, $\ct_e\approx-0.25525$, and ${\cal A}^{(t)}\approx 1.3\times 10^3$, ${\cal A}^{(v)}\approx 1.43\times 10^4$, ${\cal A}^{(s)}\approx 1.56\times 10^5$, respectively. 
  
It is worth to note the following fact. It is naturally to suppose that all modes of perturbations--scalar, vector, tensor--initially appear at moment of time $\ct_i$ with approximately equal amplitudes. However, to the end of the inflationary stage, at the time $\ct_e$, modes are amplified differently. To characterize this difference, one can introduce two parameters, $r_k=h(\ct_e,k)/\Psi(\ct_e,k)$ and $q_k=Q(\ct_e,k)/\Psi(\ct_e,k)$,
which are \textit{tensor-to-scalar} and \textit{vector-to-scalar} ratios generated during the inflationary stage for a specific wave number $k$.
%
The parameter $r_k$ can be estimated as $r_k=(\ct_e/\ct_i)^2$. Numerical results give 
$r_k\approx0.0084$. Note that the current limit from combined Planck with BICEP/Keck 2015 data \cite{Tristram:2020} yields an upper limit of an integral value $r<0.044$ at 95\%. Thus, our estimate for $r_k$ is consistent with the observable constraints. The theoretical estimate for the vector-to-scalar ratio is $q_k\approx0.091$. Here it is worth to recall that we estimate the value of $q_k$ at the end of the quasi-de Sitter (inflationary) stage and at the beginning of the post-inflationary stage. To characterize a further evolution of the vector-to-tensor ratio parameter $q_k$, one needs to take into account not only the non-minimally coupled scalar field, but also an ordinary matter filling the Universe. Anyway, in the post-inflationary stage the vector modes decay as $\ct^{-1}$ (see Eq. \rf{vmdecay}), while the long-wavelength scalar modes with $k\ct\ll 1$ are constant (see Eq. \rf{lwsmdecay}). Hence, the ratio $q_{\{k\ct\ll 1\}}\sim \ct^{-1}\sim \rma^{-2}$ decays during the post-inflationary stage. 

While the behavior of vector modes in the quasi-de Sitter (inflationary) stage is the same for \textit{all} mode wavelengths, the \textit{short-wavelength} ($k\ct\gg1$) scalar and tensor modes behave not so as {long-wavelength} ones. 
Namely, the short-wavelength scalar modes are amplified during the inflationary stage as follows:
$$
\frac{\Psi_{e,k}}{\Psi_{i,k}} \approx \frac{\Phi_{e,k}}{\Phi_{i,k}}
\approx 
\,\left(\frac{\ct_i}{\ct_e}\right)^2 e^{k(\ct_e-\ct_i)/\sqrt{3}}\equiv {\cal B}_k^{(s)}, 
$$
and the short-wavelength tensor modes as
$$
\frac{h_{e,k}}{h_{i,k}} \approx 
\,\left(\frac{\ct_i}{\ct_e}\right)^2 e^{k(\ct_e-\ct_i)}\equiv {\cal B}_k^{(t)}. 
$$
The amplifying factors ${\cal B}_k^{(s)}$ and ${\cal B}_k^{(t)}$
are now exponentially depending on the wave number $k$. This means that there exists some critical value $k_*^{(s)}=\sqrt{3}\,k_*^{(t)}$ such that all modes with $k\geq k_*^{(a)}$ ($a=\{s,t\}$) grow so intensively that
\bea
\Psi_{e,k\geq k_*^{(s)}} &=& {\cal B}_{k\geq k_*^{(s)}}^{(s)}\Psi_{i,k\geq k_*^{(s)}}\geq 1,
\nonumber\\
\Phi_{e,k\geq k_*^{(s)}} &=& {\cal B}_{k\geq k_*^{(s)}}^{(s)}\Phi_{i,k\geq k_*^{(s)}}\geq 1,
\nonumber\\
h_{e,k\geq k_*^{(t)}} &=& {\cal B}_{k\geq k_*^{(t)}}^{(t)} h_{i,k\geq k_*^{(t)}}\geq 1.
\nonumber
\eea 
In turn, this means that the linear perturbation analysis is not applicable for studying an evolution of the short-length modes with $k\geq k_*^{(a)}$ (at least in the whole inflationary stage). Instead, one has to use some nonlinear methods. It should be noted that, from a physical point of view, the transition to nonlinear evolution of scalar perturbation modes means the appearance of large-scale structures in the Universe with scales of $l\leq l_*=(k_*^{(s)})^{-1}$.
The numerical analysis performed for $\eta=0.001$ gives the following values: $k_*^{(s)}\approx 21.48$ and $k_*^{(t)}\approx 12.4$.

Finally, we would like to announce that investigations of cosmological perturbations in the theory of gravity with non-minimal derivative coupling will be continued, and our next paper which is in preparation will be devoted to constructing and analyzing a power spectrum of perturbations.

\section*{Acknowledgement}
This work is supported by the Foundation for the Development of Theoretical Physics and Mathematics “Basis,” Grant No. 24-1-1-39-1.

\end{document}